\documentclass[aip,rsi,reprint,floatfix]{revtex4-1}
\usepackage{graphicx}

\newlength{\onecolfig}
\newlength{\twocolfig}
\newcommand{\ion}[2]{\mbox{$^{#2}$#1$^+$}}
\newcommand{\Ca}[1]{\ion{Ca}{#1}}

\newcommand{\hfslev}[3]{\mbox{#1$^{\mbox{\tiny$#3$}}_{\mbox{\tiny$#2$}}$}}

\newcommand{\unit}[1]{\,\mbox{#1}}
\newcommand{\mHz}{\unit{mHz}}
\newcommand{\Hz}{\unit{Hz}}
\newcommand{\kHz}{\unit{kHz}}
\newcommand{\kHzpuT}{\unit{kHz/$\mu$T}}

\newcommand{\mH}{\unit{mH}}

\newcommand{\K}{\unit{K}}

\newcommand{\degree}{\mbox{$^{\circ}$}}

\newcommand{\ms}{\unit{ms}}
\newcommand{\us}{\unit{$\mu$s}}

\newcommand{\mA}{\unit{mA}}
\newcommand{\A}{\unit{A}}
\newcommand{\V}{\unit{V}}

\newcommand{\W}{\unit{W}}

\newcommand{\mT}{\unit{mT}}
\newcommand{\uT}{\unit{$\mu$T}}
\newcommand{\nT}{\unit{nT}}
\newcommand{\nTpK}{\unit{nT/K}}
\newcommand{\nTprtHz}{\unit{nT/$\sqrt{\mathrm{Hz}}$}}
\newcommand{\nTph}{\unit{nT/hour}}
\newcommand{\dB}{\unit{dB}}

\newcommand{\ppm}{\unit{ppm}}

\newcommand{\ish}{\mbox{$\sim$}\,}

\newcommand{\sub}[1]{\mbox{$_{\mbox{\tiny #1}}$}}
\newcommand{\diff}[1]{\mbox{\/d$#1$}}

\begin{document}

\title{Magnetic field stabilization system for atomic physics experiments}

\date{20 March 2019}

\author{B.~Merkel, K.~Thirumalai, J.~E.~Tarlton, V.~M.~Sch\"{a}fer, C.~J.~Ballance, T.~P.~Harty, D.~M.~Lucas}
\affiliation{\mbox{Department of Physics, University of Oxford, Clarendon Laboratory, Parks Road, Oxford OX1 3PU, U.K.}}

\begin{abstract}
Atomic physics experiments commonly use millitesla-scale magnetic fields to provide a quantization axis. As atomic transition frequencies depend on the magnitude of this field, many experiments require a stable absolute field. Most setups use electromagnets, which require a power supply stability not usually met by commercially available units. We demonstrate stabilization of a field of $14.6\mT$ to $4.3\nT$ rms noise ($0.29\ppm$), compared to noise of $>100\nT$ without any stabilization. The rms noise is measured using a field-dependent hyperfine transition in a single \Ca{43} ion held in a Paul trap at the centre of the magnetic field coils. For the \Ca{43} ``atomic clock'' qubit transition at $14.6\mT$, which depends on the field only in second order, this would yield a projected coherence time of many hours. Our system consists of a feedback loop and a feedforward circuit that control the current through the field coils and could easily be adapted to other field amplitudes, making it suitable for other applications such as neutral atom traps. 
\end{abstract}

\maketitle

\section{Introduction}

In many atomic physics experiments, the quantization axis is defined by a static magnetic field, typically between a few hundred microtesla and a few tens of millitesla. In trapped-ion quantum information experiments, for example, hyperfine or optical transitions can serve as the basis for qubits, when orientational degeneracy is lifted by application of a static magnetic field.\cite{Cirac1995, Steane1997, Wineland1998} The transition frequencies usually depend on magnetic field to first order, through the Zeeman effect, which leads to qubit decoherence via magnetic field noise. At certain magnetic field values, particular transitions are field-independent to first order, giving so-called ``atomic clock'' qubits; these occur for example at $11.9\mT$ in \ion{Be}{9}, $14.6\mT$ in \Ca{43}, and $21.3\mT$ in \ion{Mg}{25}.\cite{Langer2005,Harty2014,Ospelkaus2011} However, to access the qubit state, transfer pulses on field-dependent transitions are used, which makes the qubit state preparation and readout operations susceptible to magnetic field noise.\cite{Harty2014a} Hence there is a requirement for low absolute field noise at relatively large fields. Other applications, such as atomic clocks and precision tests of fundamental physics, also benefit from low magnetic field noise.\cite{Treutlein2004, Szmuk2015} 

In most setups that use magnetic coils to generate the desired field, the field noise is dominated by the coil current noise. While the use of off-the-shelf low-noise power supplies can reduce this contribution, their noise levels are still too high for high-fidelity experiments. Magnetic field noise could be minimized by using superconducting coils, which require cryogenic temperatures. While cooling the trap would benefit other properties as well, for example it should reduce the ion heating rate\cite{Niedermayr2014,Brownnutt2015} and allow for superconducting magnetic shielding\cite{Poitzsch1996,Brandl2016}, it increases the complexity of the trap design. A different approach substitutes permanent magnets for the coils, which results in a noise reduction by many orders of magnitude but loses the flexibility of adjusting the field amplitude.\cite{Sinclair2005, Ruster2016} 

Another contribution to the magnetic field noise is from the lab environment, e.g., power supplies and other electronic devices. Particularly at $50\Hz$ and harmonics, noise arising from the mains electricity is commonly observed in the laboratory. Unfortunately, environmental noise cannot be controlled directly or reduced by the use of permanent magnets. 
The most effective way to make the ions insensitive to external magnetic field fluctuations is by shielding, e.g., as provided by mu-metal enclosures,\cite{Brandl2016, Ruster2016} but this often reduces the optical access. 

Magnetic field noise can be actively cancelled by modulating the coil current to stabilize the output of a magnetometer in a feedforward or feedback configuration; possible magnetometers include fluxgates\cite{Dedman2007,Brys2005} or sense coils\cite{Marzetta1961, Glickman2012}. A limitation of this approach arises when the magnetometer cannot be located at the site of interest, for example due to physical constraints in the apparatus or saturation of the sensor. In this case, the field at the sensor may not perfectly track the field in the region of interest, leading to imperfect noise suppression.\cite{Brys2005, Dedman2007}

While the control of magnetic fields of a few hundred microtesla at a level of 2.5\ppm\ had been demonstrated using feedback and feedforward techniques,\cite{Kotler2013} the stabilisation of fields of tens of millitesla remains challenging due to the greater impact of coil current noise.\cite{FluehmannMSc} 

In this work, we present a setup to produce a 14.6\mT\ field with 0.29(1)\ppm\ rms noise. We first stabilize the 60\A\ coil current by feeding back on a low-noise current probe. Next, we suppress fluctuations in the ambient field at harmonics of the 50\Hz\ mains power frequency by adding an out-of-phase feedforward signal to the coil current, with signal parameters calibrated using a single \Ca{43} ion as a field probe. 

We characterise the performance of this system by measuring the coil current noise and the coherence time of a field-dependent qubit transition in \Ca{43}. Through a combination of feedback and feedforward, we achieve a 25-fold increase in the qubit's coherence time, consistent with the measured suppression of coil current noise.

\section{Design overview}

\begin{figure*}[htb]
\includegraphics[width=\twocolfig]{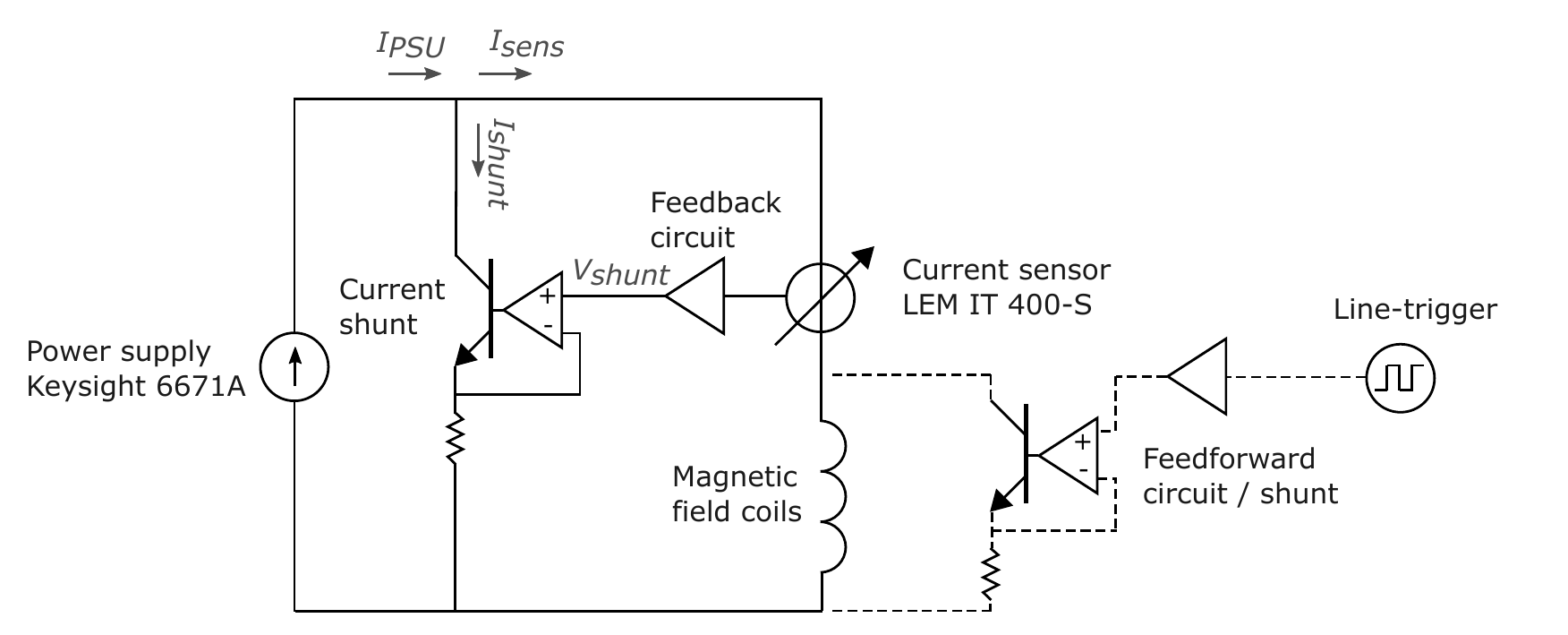}
\caption{
Schematic view of the feedback loop (solid) and the feedforward circuit (dashed). The power supply runs in constant-current mode. For the feedback loop, a sensor measures the current through the coils, from which a control voltage $V\sub{shunt}$ is derived. This control voltage is then fed into a shunt that bypasses sensor and coils. A block diagram of the feedback circuit is shown in fig.~\ref{fig:blockdiagram}. A line-trigger provides a phase-locked reference for synchronization of the feedforward correction with the mains. Full details of the circuits were published on GitHub.\cite{schematics}}
\label{fig:schematics}
\end{figure*}

Our approach to stabilizing the magnetic field has two parts: a feedback loop and a feedforward circuit. Both of them adjust the current through the magnetic field coils, which is provided by a low-noise constant current power supply \textit{Keysight~6671A} (see Fig.~\ref{fig:schematics} for the basic schematics). 

The input to the feedback loop is derived from a current sensor in series with the field coils. The sensor output is processed by several filter stages. The control voltage generated by the feedback circuit then drives the base current of a transistor that bypasses both coils and sensor, thereby adjusting the coil current and the magnetic field. Since the feedback stabilizes only the coil current and not the magnetic field itself, the ion is still susceptible to ambient magnetic field noise. We can correct for oscillations at $50\Hz$ and harmonics with a fixed phase offset from the mains electricity using the feedforward circuit, which modulates the coil current (but not the current through the sensor) in antiphase with the power line.

For the optimum feedback loop design, we start with a low-noise current sensor and design the subsequent input stage so as to minimize degradation of the sensor performance. The sensor must provide high sensitivity and low noise over a bandwidth of a few kilohertz in order to detect all noise produced by the power supply. For currents of a few amperes or less, a four-terminal resistor would be an excellent choice. At higher currents, however, there is a strong trade-off between signal level and power dissipation: a sense resistor that provides a signal of 2\V\ at 60\A\ would dissipate \ish 100\W. For this reason, we prefer to use a fluxgate sensor, which has the added benefit of being non-invasive so that the large coil current does not need to pass directly through the stabilization circuit. We chose the \textit{LEM~IT~400-S} as it was the lowest-noise sensor commercially available (specified maximum rms noise of $1\ppm$ in $0\dots1\kHz$, at $400\A$). Since its relative output noise decreases with increasing current, it is advantageous to maximize the primary current. By winding up 6 turns of the cable around the sensor, we increase the effective primary current to $360\A$. The output current of the fluxgate sensor is measured with a four-terminal resistor and an instrumentation amplifier for common-mode noise rejection. 

By subtracting the DC setpoint, we generate an error signal, i.e.\ the deviation of the sensed current from the setpoint. We measure this error signal with an FFT analyzer to obtain noise spectra. In the following section we will discuss different contributions to these noise spectra and how we generate a feedback signal from this input.

\section{Electrical characterization and feedback circuit design}
\label{S:design}

Before designing the feedback loop and the feedforward circuit, we characterize the sensor, input stage, power supply unit and coils. These measurements will determine the required feedback bandwidth and the eventual current noise floor of the closed-loop system. 

The first stages of the feedback circuit have to be designed carefully since the expected AC signals (i.e.\ noise of the coil current) are of the same order as the intrinsic noise of the components. The setpoint voltage is provided by two digital-to-analog converters (DACs) that are supplied by a low-noise voltage reference (\textit{LTC6655}). They allow for separate coarse and fine tuning of the DC setpoint, with a resolution of $0.09\ppm$ ($1.3\nT$ for our application at $14.6\mT$). In the subtraction stage, the error signal is also amplified by a gain of $200$, after which the signal is much larger than the typical noise from electronic components in the circuit, so standard design techniques can be used for the subsequent stages. 

\begin{figure}
\includegraphics[width=\onecolfig]{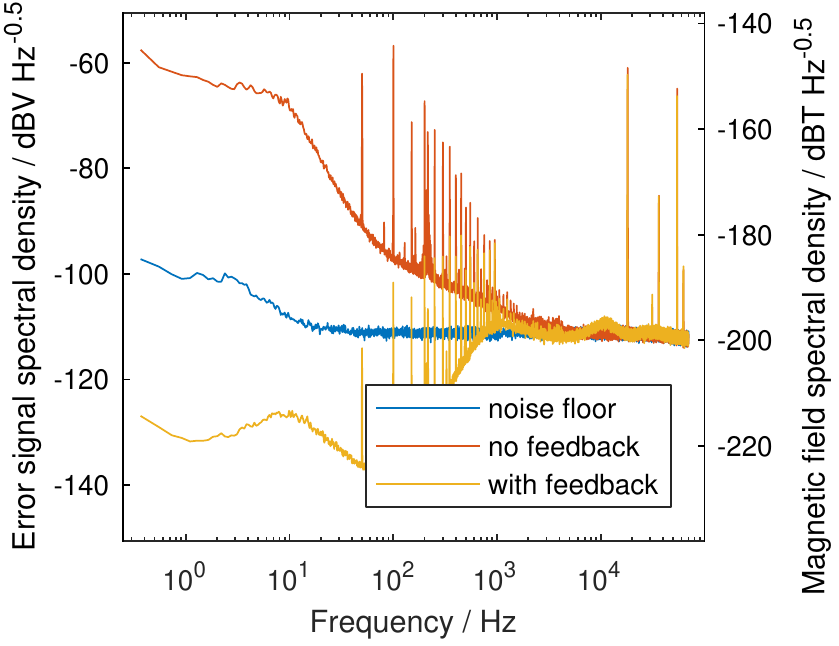}
\caption{
Measurements of the open-loop (red) and closed-loop (yellow) error signals, compared with the noise floor of the feedback loop input stage (blue). The right-hand axis gives the inferred magnetic field noise from the open-loop measurement. By closing the feedback loop, we can suppress the noise by up to $40\dB$. The closed-loop measurement appears to be below the input stage noise floor because we are measuring the in-loop error signal; the actual field noise will not be below the noise floor. 
}
\label{fig:fieldnoise}
\end{figure}

We characterize the sensor with the primary current circuit detached by measuring the error signal of the input stage with the FFT analyzer. The measured sensor noise spectrum shows white noise of $0.1\nTprtHz$ between $10\Hz$ and $80\kHz$ (see Fig.~\ref{fig:fieldnoise}, blue curve, and Table~\ref{T:currentnoise}). Spurs at $17\kHz$ and harmonics ($\sim9\ppm$ rms) are sensor artefacts arising from the fluxgate operation. As a comparison, we emulate an ideal current sensor by shorting the four-terminal resistor (i.e.\ no output current noise at $0\A$). The resulting spectrum does not show the fluxgate clock noise but has the same white noise of $0.1\nTprtHz$ ($0.26\ppm$ in $1\Hz\dots1\kHz$, equivalent to $3.8\nT$ at $14.6\mT$). This shows that the measured broadband noise level is dominated by the instrumentation amplifier or the setpoint subtracter, and not by the fluxgate sensor. Any stabilization by the feedback loop can reduce the noise no lower than this noise floor. At the low-frequency part of the spectrum, we expect noise contributions from the voltage reference and the DACs to be of the same order as the sensor's noise. Note that this measurement of the noise floor was taken at zero DC setpoint, where the reference voltage noise does not contribute to the analyzed signal. Hence, the measured noise floor is only a lower bound at low frequencies. For a more detailed analysis of the noise contributions from different parts of the circuit, we provide a noise and stability model online along with the source files.\cite{SupplInf} 

We next connect the sensor to the coils and the power supply unit, without the feedback or feedforward shunts, and apply a current of $60\A$ to measure the initial coil current noise without any stabilization (fig.~\ref{fig:fieldnoise}, red curve). The noise of the power supply is mainly located at frequencies below $5\kHz$, leading to $96\nT$ rms noise between $0.4\Hz$ and $1\kHz$. Spurs at $50\Hz$ and harmonics ($>50\nT$ rms) are fluctuations of the coil current with the mains electricity. 

From these pre-characterization measurements we derive a required bandwidth of the feedback loop between $1\kHz$ and $10\kHz$: on one hand, the bandwidth should be no lower than $1\kHz$ since otherwise a major part of the power supply noise would not be attenuated. As the bandwidth increases, the attenuation factor at lower frequencies increases, too. On the other hand, the feedback must not respond to the fluxgate clock noise at $17\kHz$ and harmonics. For this reason, we add notch and low-pass filters to the feedback circuit after the error-signal generation. These filters, however, reduce the phase margin of the feedback loop, which could result in instability if the bandwidth was close to the notch frequency ($17\kHz$). Therefore, a trade-off has to be found between the suppression of any $17\kHz$ fluxgate clock noise, and the suppression of broadband noise. We choose $3\kHz$ as the target bandwidth. 

We now have to ensure correct operation of the current shunt over the target bandwidth of the feedback loop. If the power supply were an ideal constant-current source with a DC output current $I\sub{PSU}$, any AC signal $I\sub{shunt}$ through the shunt would induce the same signal but with opposite sign in the current $I\sub{sens}$ through sensor and magnetic field coils, since $I\sub{sens}=I\sub{PSU}-I\sub{shunt}$. The supply's large output capacitance, however, leads to a deviation from an ideal constant current source at higher frequencies and gives resonant effects by coupling to the coils' inductance. We characterize this behaviour by measuring the transfer function $I\sub{sens}/I\sub{shunt}$, which shows a drop in amplitude of nearly two orders of magnitude from $10\Hz$ to $100\Hz$. To correct for this, we add a compensation stage after the notch filters of the feedback circuit. We find that a bi-quadratic filter works well for our setup, with components chosen so that the filter's transfer function approximates the inverse of the measured one. Note that the measured transfer function depends strongly on the power supply (and also on the coil inductance, which is $\approx 0.6\mH$ in our experiment), and therefore the components in the compensation stage would have to be adapted to different setups. 

\begin{figure*}[htb]
\includegraphics[width=\twocolfig]{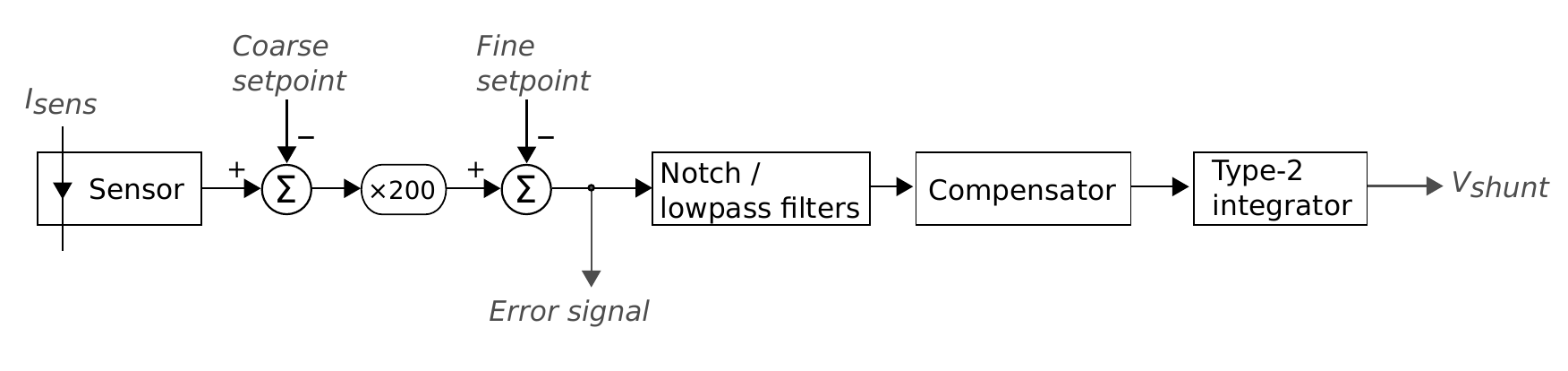}
\caption{
Block diagram of the feedback circuit. $I\sub{sens}=I\sub{PSU}-I\sub{shunt}$ is the current through the sensor and the magnetic field coils. The error signal is the sensor output after amplification and setpoint subtraction. The feedback circuit further consists of notch and low-pass filters to suppress the fluxgate clock noise, a compensator to correct for the large output capacitance of the power supply, and a type-2 integrator that provides gain. $V\sub{shunt}$ is the output signal to the shunt.}
\label{fig:blockdiagram}
\end{figure*}

The final stage of the feedback circuit provides gain and defines the bandwidth; its output is taken to the current shunt, which can shunt up to 100\mA\ away from the coils. We use a double integrator to reduce low-frequency noise without degrading the phase margin near the unity-gain cross-over. A  block diagram of the feedback circuit is shown in fig.~\ref{fig:blockdiagram}.

\begin{table}
\caption{Summary of rms magnetic field noise under various conditions. Values in parentheses are derived from measurements of the error signal with an FFT analyzer (see fig.~\ref{fig:fieldnoise}); others are deduced from measurements on the \Ca{43} qubit (see table~\ref{tab:coherencetimes}). }
\begin{ruledtabular}
\begin{tabular}{llp{3.8cm}}
   & rms noise & dominant noise source \\ \hline
   No stabilization   & ($96\nT$) & power supply unit \\
   Feedback alone     & $16.1\nT$   & ambient 50\Hz\ line noise \\
   With feedback      & $4.5\nT$    & limited by noise floor and \\ 
   ~~and feedforward    &          & ~~feedback bandwidth \\
   Sensor noise floor & ($3.8\nT$)  & instrumentation amplifier,\\
   ~~(0.1\Hz\ to 3\kHz) &			      & ~~reference subtraction \\
\end{tabular}
\end{ruledtabular}
\label{T:currentnoise}
\end{table}

For initial characterization, we measure the current noise with the feedback loop closed (Fig.~\ref{fig:fieldnoise}, yellow curve) and compare it to the previous measurements without any stabilization (red curve). When the feedback loop is closed, the noise is clearly suppressed for frequencies below $3\kHz$. It appears to be below the noise floor because we are measuring the in-loop error signal; the actual current noise will not be lower than the noise floor as determined before (blue curve). Around $10\kHz$, a small bump in the noise spectrum is only present with feedback enabled: it arises where the phase margin is $<90\degree$ as a result of the trade-off chosen between the cancellation of any $17\kHz$ fluxgate clock noise and the suppression of broadband noise around $10\kHz$. 

\section{Feedforward calibration}

Since the ion is sensitive to the magnetic field and not just to the coil current, we still have contributions from ambient field noise. A dominant contribution to ambient noise arises from the mains electricity, at 50\Hz\ and harmonics thereof. While the feedback loop suppresses current noise at these frequencies sufficiently, the ambient field noise is not attenuated, as it is not detected by the current sensor. 

To probe the magnetic field, we use a single trapped \Ca{43} ion that is held in a blade-type Paul trap in ultrahigh vacuum and is located approximately at the geometrical centre of the magnetic field coils.\cite{SarahWoodrowMSc} We perform Ramsey experiments on a qubit stored in the ground level hyperfine \hfslev{4S}{1/2}{4,+4} and \hfslev{4S}{1/2}{3,+3} states, whose frequency splitting is first-order dependent on the magnetic field via the Zeeman effect, with coefficient 24.5\kHzpuT. Magnetic field fluctuations will shift the qubit frequency and translate into a phase shift accumulated in each measurement shot. 

To measure the effects of magnetic field modulation in phase with the 50\Hz\ power line cycle, we use the zero-crossing of the power line cycle as a trigger for our experiments. By running a Ramsey sequence much shorter than $20\ms$ and analysing the phase accumulated by the qubit, we can determine the qubit frequency for a fixed delay time from the line-trigger. Since the qubit frequency follows the magnetic field, a series of measurements for different delay times therefore reveals the oscillations of the magnetic field at the ion within each mains electricity cycle. 

\begin{figure}
\includegraphics[width=\onecolfig]{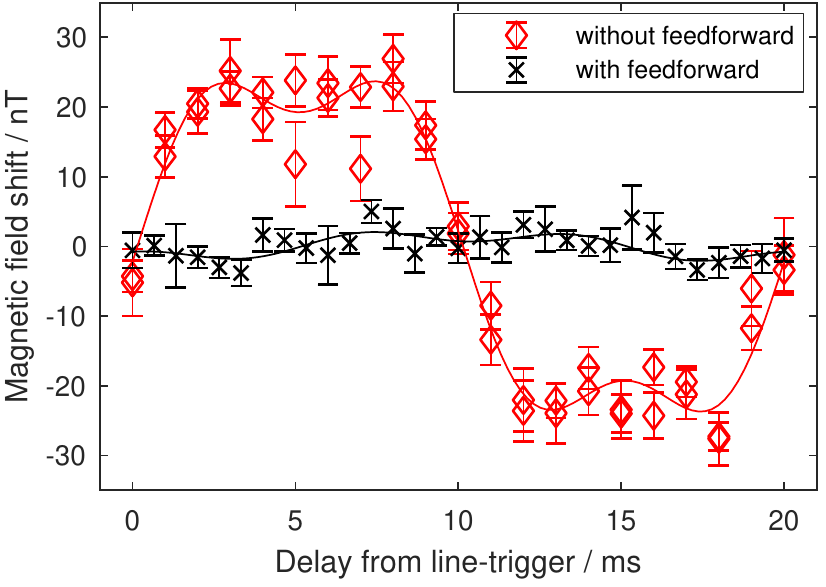}
\caption{
Oscillation of the magnetic field within each 20\ms\ mains cycle, deduced from the $\hfslev{4S}{1/2}{4,+4}\leftrightarrow\hfslev{4S}{1/2}{3,+3}$ transition frequency of the \Ca{43} qubit, which was measured with a $\tau_R=10\us$ Ramsey experiment. Error bars are statistical. The modulation without (with) the feedforward circuit is shown in red (black). All experiments were line-triggered, and the feedback was enabled for both datasets. The solid curves are fits to modulation with components at $50\Hz$ and $150\Hz$; the amplitude coefficients without (with) feedforward are $26.2(5)\nT$ and $7.0(8)\nT$ ($1.7(4)\nT$ and $1.0(4)\nT$). 
}
\label{fig:linecycle}
\end{figure}

With the feedback loop enabled, we measure a $50\Hz$ modulation with an amplitude of $26.2(5)\nT$ (Fig.~\ref{fig:linecycle}, red dataset) (for comparison, the current noise at $50\Hz$ and harmonics without any stabilization corresponds to $50\nT$ rms). 

This coherent modulation is clearly distinguishable from broadband noise and fortunately does not change significantly over periods of weeks. This allows us to program the feedforward circuit to modulate the coil current in (anti)phase with the power line to counteract the magnetic field oscillation measured at the ion. With this feedforward circuit, we measure a remaining modulation by the mains electricity of only $1.4(5)\nT$ rms, limited by the measurement accuracy (Fig.~\ref{fig:linecycle}, black dataset).

\section{Coherence times and long-term stability}

As a figure of merit for the overall magnetic field stability, we measure the qubit coherence time by Ramsey experiments. Due to magnetic field noise, the accumulated phase differs from measurement shot to shot, resulting in a drop of the Ramsey fringe contrast $c$ when averaged over many shots. In our measurements we observe a Gaussian decay $c=\exp(-\tau_R^2/T_c^2)$, which indicates correlated noise on a time scale longer than the Ramsey delay time $\tau_R$. The magnetic field rms noise $\sigma_B$ is related to the Ramsey coherence time $T_c$ by\cite{OMalley2015,Brouard2003} 
\begin{equation}
\label{eq:Bnoise-Tc}
\sigma_B=\frac{\sqrt{2}}{T_c\;(2\pi\times 24.5\kHzpuT)}.
\end{equation}

Before stabilization and without line-triggering, we measure a coherence time of $0.082(2)\ms$, which is in good agreement with a simulation based on the measured current noise spectrum.\cite{Uhrig2007,Szwer2011} With the feedback loop enabled, we extend the coherence time to $0.57(1)\ms$, implying a remaining rms field noise of $16.1(3)\nT$ (Table~\ref{tab:coherencetimes}). This is consistent with the measured mains field, which produces a random variation in the field for each shot. 

A common way to avoid the dephasing by a field modulation at $50\Hz$ and harmonics is to synchronise the experiments with the mains cycle so that the field is the same for each shot of the experiment.\cite{SchmidtKaler2003} In such line-triggered Ramsey experiments (0\ms\ delay in Fig.~\ref{fig:linecycle}) we measure a coherence time of $1.93(7)\ms$, which translates to a magnetic field noise of $\sigma_B=4.8(2)\nT$. 

Alternatively, we can use the feedforward circuit to suppress the field oscillations at $50\Hz$ and harmonics. The fully-stabilized experiments with both feedback and feedforward enabled show a coherence time of $2.03(5)\ms$ (rms field noise of $4.5(1)\nT$), which is consistent with the result obtained from line-triggered experiments without a feedforward circuit (see Table~\ref{tab:coherencetimes} for comparison). Thus the feedforward circuit eliminates the need for line-triggering, which can greatly improve the rate of data acquisition, as the experimental repetition rate is not limited to 50\Hz.

The observed Ramsey coherence time of about 2\ms\ is in agreement with a simulation based on the noise floor from the current noise measurements (Fig.~\ref{fig:fieldnoise}, Table~\ref{T:currentnoise}) after correcting it for missing low-frequency noise that might be added by the reference.\cite{Uhrig2007, SupplInf} Thus, the achieved stabilization is as good as one could expect from the specified noise levels of the circuit's components. 

\begin{figure}
\includegraphics[width=\onecolfig]{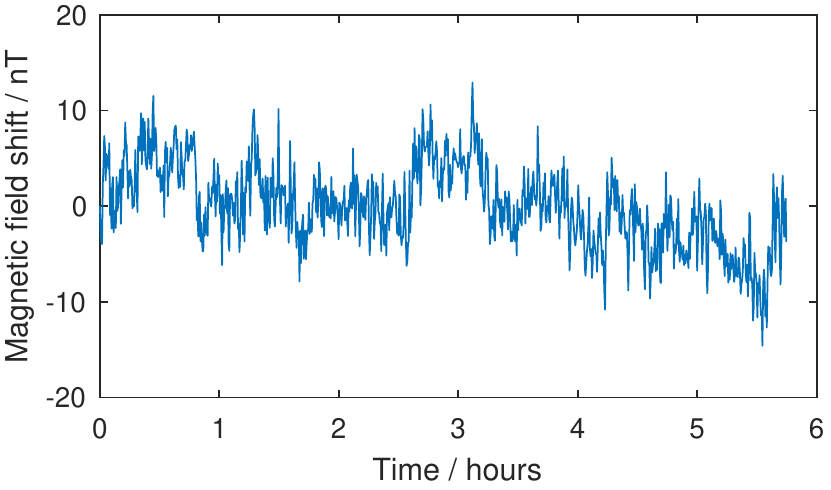}
\caption{
Long-term magnetic field fluctuations measured at the ion, by monitoring the 
$\hfslev{4S}{1/2}{4,+4}\leftrightarrow\hfslev{4S}{1/2}{3,+3}$ qubit frequency over a period of several hours. The feedback loop was closed. The rms noise on an hour time scale is $4.3\nT$, and the average drift about $1.5\nTph$. 
}
\label{fig:longtermstability}
\end{figure}

In spin-echo measurements, which are less sensitive to noise at frequencies below $\ish 2/\tau_R$, the observed coherence times are about five times longer compared with simple Ramsey sequences (see Table~\ref{tab:coherencetimes}). This indicates that the main source of decoherence is low-frequency noise. We investigate this by monitoring the drift of the qubit frequency on an hour timescale (see Fig.~\ref{fig:longtermstability}). With the feedback enabled, the slow drift corresponds to an rms noise of $4.3\nT$, or 0.29ppm (for frequencies $<0.05\Hz$), which is similar to the results obtained from the coherence time measurements without spin-echo. 
This slow drift is consistent with the observed level of $0.29\ppm$ low-frequency rms noise to the fluxgate sensor, the DACs and the voltage reference used in the feedback loop. Note that we do not expect temperature to play a significant role in long-term stability as the temperature coefficient of the circuit is estimated to be $21\nTpK$ (dominated by the voltage reference and the fluxgate sensor) and the the circuit was operated in an environment stable to $0.1\K$ during these measurements.\cite{SupplInf} 

For comparison, without any stabilization, the noise was predominantly at frequencies below $10\Hz$ (see Fig.~\ref{fig:fieldnoise}), with a low-frequency ($<1\Hz$) rms noise of about $14\ppm$. 

\begin{table}
\caption{\label{tab:coherencetimes}Qubit coherence times from Ramsey and spin-echo measurements, and magnetic field noise as deduced using eq.\ref{eq:Bnoise-Tc}. The noise values for the spin-echo experiments correspond to a high-pass filtered noise spectrum.\cite{Biercuk2010, Uhrig2007} FB indicates experiments performed with feedback enabled; FF with feedforward enabled; LT line-triggered experiments synchronized with the 50\Hz\ mains cycle.}
\begin{ruledtabular}
\begin{tabular}{l|rr|rr}
stabilization~ & \multicolumn{2}{c|}{coherence time} & \multicolumn{2}{c}{magnetic field noise} \\
  & \multicolumn{2}{c|}{$T_c$ (ms)} & \multicolumn{2}{c}{$\sigma_B$ (nT rms)} \\ \hline 
  & Ramsey & spin-echo~ & Ramsey & spin-echo \\
	none & $0.082(2)$ &  & $112(3)$ & \\
	FB & $0.57(1)$ &  & $16.1(3)$ & \\
	FB, LT & $1.93(7)$ & $11.3(2)$ & $4.8(2)$ & $0.81(1)$ \\
	FB, FF & $2.03(5)$ & $9.6(1)$ & $4.5(1)$ & $0.96(1)$ \\
\end{tabular}
\end{ruledtabular}
\end{table}

\section{Summary}

We have developed feedback and feedforward circuits that can stabilize a magnetic field of $14.6\mT$ to a noise level of $4.3\nT$ rms ($0.29\ppm$). This is sufficient to give a projected coherence time of many hours for the \Ca{43} atomic clock qubit at 14.6\mT\ (which has a second-order field-dependence of $\diff{^2 f}/\diff{B^2} = 240\mHz/\uT^2$). The performance of the stabilization circuits is not restricted to this particular field strength and they can be straightforwardly adapted to other experiments with different magnetic field requirements. To obtain a similar relative noise level of the sensing stage, the number of turns for which the coil power cable is wrapped around the fluxgate sensor needs to be adjusted. All other corrections specific to the coils and the power supply are implemented by the compensator stage of the feedback circuit, which can determined by a measurement of the transfer function as described in section~\ref{S:design}.  An alternative strategy would be to replace the analogue feedback circuit used here with a digital signal processing (DSP) unit, with which the feedback optimization could be automated.

The coherence times that we measured for the field-dependent qubit in \Ca{43} at a stabilized field of $14.6\mT$ are comparable to those obtained for field-dependent qubits at low magnetic fields without stabilization (e.g.\ $11(2)\ms$ in line-triggered spin-echo measurements on a \Ca{40} Zeeman qubit at 0.37\mT)\cite{Ruster2016}, indicating that we have achieved similar absolute field stability at 14.6\mT\ as is typically obtained at fields below 1\mT. This is also advantageous for experiments using multiple ion species,\cite{Ballance2015,Tan2015} as the different species do not possess atomic clock transitions at the same magnetic field.

\section*{Supplementary Material}

A detailed analysis of the broadband and low-frequency noise and the long-term stability of the components used in the stabilization circuit has been published as supplementary material, as well as schematics of the feedback circuit and the current shunt.  

\section*{Acknowledgements}

We thank D.~P.~Nadlinger for comments on the manuscript. J.E.T.\ acknowledges funding from the Centre for Doctoral Training on Controlled Quantum Dynamics at Imperial College London. This work was supported by the U.S.\ Army Research Office (ref.\ W911NF-14-1-0217) and the U.K.\ EPSRC ``Networked Quantum Information Technology'' Hub.

\bibliography{StabiliserBibliography}


\end{document}